\documentstyle[aps,pra,epsfig,manuscript,eqsecnum]{revtex} \input epsf

\begin{document}

\draft
\title{Finite Temperature Collapse of a Bose Gas with Attractive
  Interactions}
\author{ Erich J.~Mueller\cite{REF:EJM} and Gordon Baym\cite{REF:GB}}
\address{ Department of Physics, University of Illinois at
  Urbana-Champaign, 1110 West Green Street, Urbana, IL 61801}
\date{May 16, 2000}
\maketitle
\begin{abstract}
  We study the mechanical stability of the weakly interacting Bose gas
  with attractive interactions, and construct a unified picture of the
  collapse valid from the low temperature condensed regime to the high
  temperature classical regime.  As we show, the non-condensed
  particles play a crucial role in determining the region of
  stability, even providing a mechanism for collapse in the
  non-condensed cloud.  Furthermore, we demonstrate that the
  mechanical instability prevents BCS-type ``pairing'' in the
  attractive Bose gas.  We extend our results to describe domain
  formation in spinor condensates.
\end{abstract}
%
%
%

\narrowtext

\section{introduction}

The trapped Bose gas with attractive interactions is a novel physical
system.  At high densities such clouds are mechanically unstable;
however at low densities they can be stabilized by quantum mechanical
and entropic effects.  Stability and collapse has been observed in
clouds of degenerate $^7$Li \cite{REF:randy} and $^{85}$Rb
\cite{REF:JILA}.  The collapse of the Bose gas shares many features of
the gravitational collapse of cold interstellar hydrogen, described by
the Jeans instability \cite{REF:jeans}; related instabilities occur in
supercooled vapors.  Theoretical studies of the attractive Bose gas,
typically numerical, have been limited to zero \cite{REF:zeroT} or
very low temperature \cite{REF:finiteT,REF:devreese}.  Here we give a
simple analytic description of the region of stability and threshold
for collapse valid from zero temperature to well above the Bose
condensation transition, and thus provide a consistent global picture
of the instability.  As we see, at finite temperature, the phase
diagram of the Bose gas includes regions where the non-condensed
particles play a significant role in the collapse \cite{noncond}.

Our results are summarized in the phase diagram in
Fig.~\ref{FIG:phase} which shows three regions: normal, Bose
condensed, and collapsed.  This third region is not readily accessible
experimentally; the system becomes unstable at the boundary (the solid
line in the figure).  In this figure the condensation and collapse
lines actually meet at a finite angle.  The gross features of the
phase diagram can be understood qualitatively with dimensional
arguments.  At low temperatures the only stabilizing force is the
zero-point motion of the atoms.  This ``quantum pressure'' has a
characteristic energy per particle of $E_Q\sim \hbar^2 V^{-2/3}/m$,
where $\hbar$ is Planck's constant, $V$ is the volume in which the
cloud is confined, and $m$ is the mass of an atom.  The attractive
interactions which drive the collapse are associated with an energy
$U=\hbar^2 a_s n/m$, where $n$ is the density, and $a_s$ ($=-1.45\,$nm
for $^7$Li and $-20\,$nm for $^{85}$Rb) is the s-wave scattering
length.  Comparison of $U$ with $E_Q$ suggests that for low
temperatures the density $n$ at collapse should be $\sim (V^{2/3}
a_s)^{-1}$, independent of temperature.  At high temperatures, the
stabilizing force is thermal pressure, $P=-T\partial S/\partial V$,
which is characterized by the thermal energy $E_T\sim k_B T$.  Here
comparison of $U$ with $E_T$ suggests that at high temperatures the
density of collapse should be $\sim m k_B T/\hbar^2 a_s$, linear in
temperature.  The crossover between the quantum and classical
behaviors ($n\propto T^0$ and $n\propto T$) occurs near $T_c$.  The
collapse, as described here, is a phenomena in which the cloud as a
whole participates, not just the condensate.

Our basic approach is to identify the collapse with an instability in
the lowest energy mode of the system (the breathing mode in a
spherically symmetric cloud).  As the system goes from stable to
unstable, the frequency of this mode goes from real to complex,
passing through zero when the instability sets in.  The density
response function, $\chi(k,\omega)$, which measures the response of
the cloud to a probe at wave vector $k$ and frequency $\omega$,
diverges at the resonant frequencies of the cloud.  Therefore, by
virtue of the vanishing frequency of the lowest energy mode, the
collapse is characterized by
\begin{equation}\label{res}
\chi(k=2\pi/L,\omega=0)^{-1}=0.
\end{equation}
Here $k=2\pi/L$ is the wave-vector of the unstable mode, whose
wavelength $L$ should be of order the size of the system.  Equation
(\ref{res}) implicitly determines the line of collapse in the
temperature-density plane.

To evaluate the response function analytically, we use a local-density
approximation, replacing the response function $\chi(k,\omega)$ of the
inhomogeneous cloud by that of a gas with uniform density $n$.  The
response of the uniform gas is evaluated at the same frequency and
wave-vector as for the inhomogeneous system, and $n$ is given by the
central density of the atomic cloud.  The local-density approximation
should be valid for temperatures large enough that the thermal
wavelength, $\Lambda_T=(2\pi\hbar^2/m k_B T)^{1/2}$, is much smaller
than the size of the trap.  In all experiments to date, this condition
is satisfied, and we treat $k\Lambda_T$ as a small parameter in our
calculation.  With this approximation, we calculate the line of
collapse using the well developed theory of $\chi(k,\omega)$ of a
uniform gas (reviewed in \cite{REF:griffin}).

In the experiments on $^7$Li, the atoms are held in a magnetic trap
with a harmonic confining potential $V(r)=\frac{1}{2} m \omega^2 r^2$,
with $\omega\approx 2\pi\times 145 s^{-1}$\cite{REF:freq}.  This
potential gives the cloud a roughly Gaussian density profile (see
Sec.~\ref{SEC:profile}).  The temperature $T$ is typically 50 times
the trap energy $\hbar\omega\approx 7$nK, so the parameter
$k\Lambda_T$, $\approx \sqrt{\hbar\omega/k_BT} \approx 0.14$, is
small.  The experiments on $^{85}$Rb use softer traps,
$\hbar\omega\approx 0.6$nK, and colder temperatures $T\approx 15$nK,
so that $k\Lambda_T\approx 0.2$.

The goal of our approach is to provide a framework for investigating
the interplay of condensation and collapse.  Although our results are
not as accurate as can be obtained numerically (comparing with
previous numerical work \cite{REF:finiteT}, we find that are results
are always well within a factor of two of those calculated using more
sophisticated models), the conceptual and computational advantages of
working with a uniform geometry far outway any loss in accuracy.  Due
to their simplicity, the arguments used here provide an essential tool
to choosing which parameter ranges to investigate in future
experiments and computations.

The tools introduced to discuss the collapse of an attractive gas can
also be used to describe other instabilities in trapped Bose gases.
In Section~\ref{SEC:pairing} and \ref{SEC:spins}, we apply these
methods to the problem of BCS type pairing between bosons, and towards
domain formation in spinor condensates.

\section{Simple Limits}
\subsection{Zero Temperature}

To illustrate our approach we first consider the stability of a zero
temperature Bose condensate.  The excitation spectrum of a uniform gas
is \cite{REF:bog}:
\begin{equation}\label{EQ:bog}
    \omega^2 = \left( \frac {k^2} {2m} \right)^2 +  g n \frac {k^2}{m} \equiv
   E_k^2,
\end{equation}
where $g=4 \pi\hbar^2a_s/m$.  In the attractive case, $g<0$, all long
wavelength modes with $k^2/2m < 2|g|n$ have imaginary frequencies and
are unstable.  A system of finite size $L$ only has modes with
$k>2\pi/L$, and for larger $k$ has an excitation spectrum similar to
Eq.~(\ref{EQ:bog}).  If $|g|n<\hbar^2\pi^2/mL^2$, the unstable modes
are inaccessible and the attractive Bose gas is stable.

This information is included in the density response function
$\chi(k,\omega)$ of the dilute zero-temperature gas
\cite{REF:griffin},
\begin{equation}
\chi(k,\omega) = \frac{2n\varepsilon_k}{\omega^2-E_k^2},
\label{EQ:zsus}
\end{equation}
where $\varepsilon_k = k^2/2m.$ The poles of $\chi$ are at the
excitation energies, $\pm E_k$.  In particular,
$\chi(k=2\pi/L,\omega=0)$ diverges when $|g|n=\hbar^2\pi^2/mL^2$.

\subsection{High Temperature ($T\gg T_c$)}\label{SEC:hit}

We further illustrate our procedure by calculating the stability of an
attractive Bose gas at temperature much larger than $T_c$, where
thermal pressure is the predominant stabilizing force.  Quantum
effects are negligible in this limit, and the line of collapse is
simply the spinodal line of the classical liquid-gas phase transition
\cite{REF:spinodal}, as characterized by Mermin \cite{REF:mermin}.  We
neglect finite size effects, and look for an instability in the
uniform gas at zero wavevector, $k=0$, corresponding to finding where
$\chi(k=0,\omega=0)$ diverges.  The susceptibility
$\chi(0,0)=-\partial n/\partial \mu$ (where $\mu$ is the chemical
potential), is proportional to the compressability of the system,
which diverges when the gas becomes unstable.

At high temperature we work in the Hartree-Fock approximation, where
the density is given by the self-consistent solution of
\begin{equation}\label{EQ:density}
  n = \int\frac{d^3p}{(2\pi\hbar)^3}
    \frac{1}{e^{\beta (\epsilon_p -\mu)}-1},
\end{equation}
with Hartree-Fock quasiparticle energies $\epsilon_p = \varepsilon_p
+2gn$; here $\beta = 1/k_BT$.  In the classical limit ($\beta\mu\ll
-1$), $n= e^{\beta(\mu- 2g n)}/\Lambda_T^3$.  The response $\chi(0,0)$
has the structure of the random phase approximation (RPA),
\begin{equation}\label{EQ:hf}
 \chi(k,\omega) = \frac{\chi_0(k,\omega)}{1- 2g \chi_0(k,\omega)},
\end{equation}
where $\chi_0(0,0)= -\left(\partial n/ \partial\mu\right)_\epsilon$
(where the $\epsilon$ are held fixed) is the ``bare'' response.  In
the classical limit $\chi_0(0,0)=-\beta n$.  Since $\chi_0(0,0)$ is
negative, the repulsive system ($g>0$) is stable.  However, for
attractive interactions $g<0$), the denominator of Eq.  (\ref{EQ:hf})
vanishes when $2 g \chi_0 = 1$, which in the classical limit occurs
when $2|g| n=k_BT$.

    The above calculation is only valid well above $T_c$.  When
$|\mu|\sim\hbar^2/mL^2$, finite size effects start to become important, and a
more sophisticated approach is needed.  If one blindly used the above result
near $T_c$ one would erroneously find that the instability towards collapse
prevents Bose condensation from occurring.  This difficulty can be avoided by
working with the finite wave-vector response $\chi(k=2\pi/L,\omega=0)$, to
which we now turn.

\section{Density Response Function}

    The approximation we shall use for $\chi$ is to treat the response of the
gas in the RPA, with the simplifying assumption that the bare response of the
condensed and non-condensed particles is taken to be the response of a
non-interacting system.  This approach, employed by Sz\'epfalusy and Kondor
\cite{REF:szef} in studying the critical behavior of collective modes of a
Bose gas, and later modified by Minguzzi and Tosi \cite{REF:minguzzi} to
include exchange, is simple to evaluate analytically, and is valid both above
and below $T_c$.  It generates an excitation spectrum which is conserving
\cite{REF:cons} and gapless \cite{REF:griffin}.  At zero temperature it yields
the Bogoliubov spectrum, Eq.~(\ref{EQ:bog}), and above $T_c$ it
becomes the standard RPA with exchange.

\begin{figure}
  \epsfxsize=\columnwidth\nobreak
  \centerline{\epsfbox{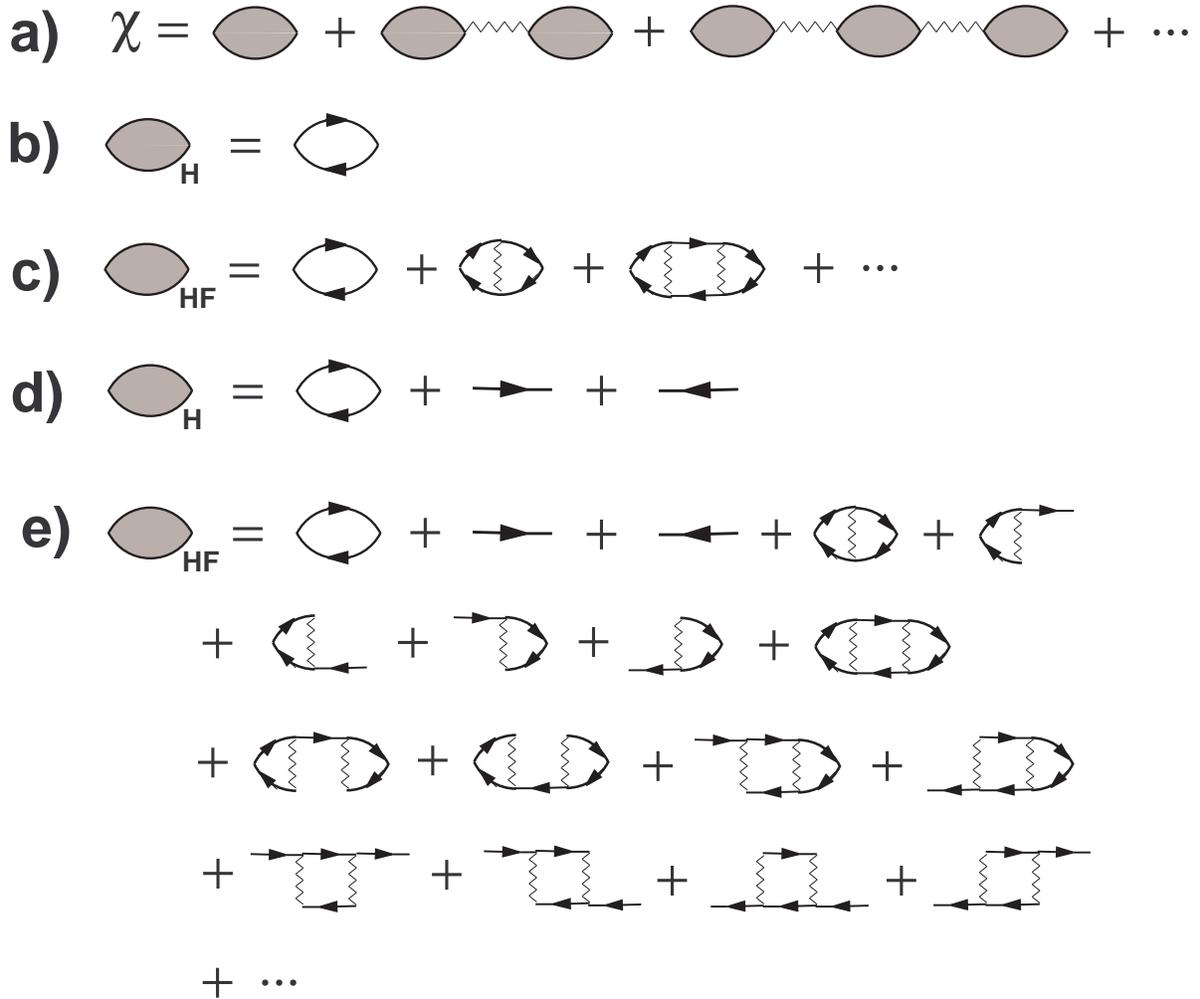}}\nobreak
\caption{
  Diagrammatic description of the approximation used for the response
  function.  Lines with arrows represent propagators of the
  non-interacting system.  Jagged lines represent the contact
  interaction between particles.  a) Generically the response function
  can be expressed in terms of a polarization bubble (the bubble with
  shaded interior).  b) In the Hartree approximation (RPA), the
  polarization bubble is taken to be the response of the
  non-interacting gas.  c) In the Hartree-Fock approximation, the
  polarization bubble involves a sum over repeated interactions.  d)
  In a condensed system, the polarization bubble for the Hartree
  approximation must include terms where one of the particles is in
  the condensate.  e) The Hartree-Fock approximation for the condensed
  gas requires writing all exchange graphs.  This series is most
  easily summed by the matrix approach in Appendix~A.}
\label{FIG:hfdiagrams}
\end{figure}

    The susceptibility in this approximation has the form,
\begin{equation}\label{EQ:sus}
\chi(k,\omega) = \frac{\chi_0^c + \chi_0^n + g \chi_0^c\chi_0^n}
{(1-g\chi_0^c)(1-2g\chi_0^n)-4g^2\chi_0^c\chi_0^n},
\end{equation}
where $\chi_0^c$ and $\chi_0^n$ are the condensate and non-condensed
particle contributions to the response of the non-interacting cloud,
\begin{mathletters}
\begin{eqnarray}\label{EQ:chio}
\chi_0^n(k,\omega) &=& \int\frac{d^3q}{(2\pi)^3}
 \frac{ f(q-k/2)-f(q+k/2)}
      {\omega-(\varepsilon_{q+k/2} -\varepsilon_{q-k/2})},\\
\label{EQ:chic}
\chi_0^c(k,\omega) &=&
  \frac{n_0}{\omega-\varepsilon_k}-\frac{n_0}{\omega+\varepsilon_k}.
\end{eqnarray}
\end{mathletters}
Here $n_0$ is the condensate density, the $\varepsilon_k=k^2/2m$ are the
free particle kinetic energies as before, and the Bose factors $f(k)$ are
given by $(e^{-\beta(\varepsilon_k-\mu)}-1)^{-1}$.  In Appendix A we briefly
review the derivation of this response function.  At zero temperature
$\chi_0^n=0$ and the susceptibility reduces Eq.~(\ref{EQ:zsus}),
while above $T_c$, $\chi_0^c=0$, and $\chi$ reduces to Eq.~(\ref{EQ:hf}).
Figure \ref{FIG:hfdiagrams} shows the class of diagrams summed in this
approximation.

\begin{figure}[tb]
  \epsfxsize=5in\nobreak
  \centerline{\epsfbox{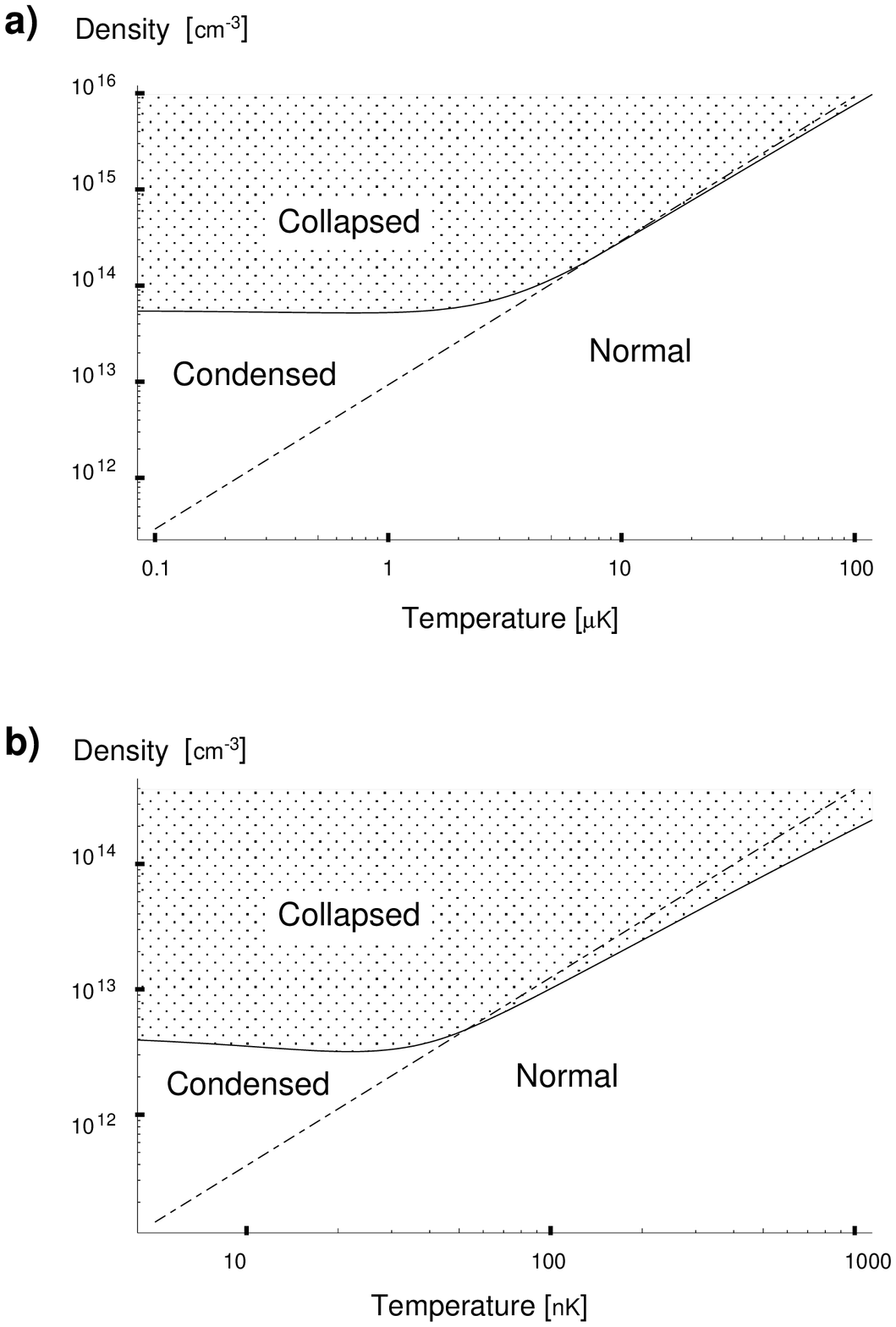}}\nobreak \vskip2mm\nobreak
 \caption{
   Phase diagram, density $n$ [cm$^{-3}$] versus temperature $T$,
   for a cloud of attractive bosons. a) For a cloud of $^7$Li with
   spatial extent $L=3.15 \mu$m, scattering length $a_s = -1.45$nm,
   corresponding to the experiments in Ref.  [1], performed at
   temperatures $T$ near $300$ nK.  b) For a cloud of $^{85}$Rb with
   scattering length $a_s = -20$nm and spatial extent $L=3 \mu$m,
   corresponding to the recent experiments
   in Ref.  [2], performed at $T$ near 15 nK.\\
   Note the logarithmic scales.  The solid line separates the unstable
   (shaded) region from the stable region.  The dashed line,
   representing the Bose condensation transition has been continued
   into the collapsed region to illustrate that the two lines
   intersect.  This diagram is drawn for a uniform but finite cloud,
   but can be applied to harmonically trapped gases by taking $n$ to
   be the central density in the trap.}
\label{FIG:phase}
\end{figure}

    Expanding $\chi_0^n$ in the small parameter $k\Lambda_T$ (see details in
Appendix~B), we derive for $T>T_c$,
\begin{equation}\label{EQ:aboveT}
g\chi_0^n(k,0) =  -2\frac{a_s}{\Lambda_T}
    \left[\frac{4\pi}{k\Lambda_T}
          \arctan\left|{\epsilon_k}/{4\mu}\right|^{1/2}
+ g_{1/2}(e^{\beta\mu})
-\left|{\pi}/{\beta\mu}\right|^{1/2}
+{\cal O}(k\Lambda_T)
\right],
\end{equation}
where $g_\nu(z) \equiv \sum_j z^j/j^\nu$ is the polylogarithm function.
For chemical potential $\mu$ much larger in magnitude than $k_B T$, the system
is classical, and Eq.~(\ref{EQ:aboveT}) reduces to $g\chi_0^n=-\beta g n$, as
in the Hartree-Fock approach, Sec.~\ref{SEC:hit}.  Below $T_c$ the chemical
potential of the non-interacting system vanishes and the response functions
are:
\begin{mathletters}
\begin{eqnarray}\label{EQ:belowT}
g\chi_0^n(k,0) &=& -\frac{4\pi^2 a_s}{k \Lambda_T^2}
+ {\cal O}((k\Lambda_T)^{0}),\\\label{EQ:belowT2}
g\chi_0^c(k,0) &=& - 16\pi\frac{a_s n_0}{k^2}.
\end{eqnarray}
\end{mathletters}
Using these expressions we calculate the spinodal line separating the
stable and unstable regions of Fig.~\ref{FIG:phase} by setting
$k=2\pi/L$ and solving the equation
\begin{equation}
 \chi^{-1}\propto 1-g(\chi_0^c+2\chi_0^n)-2 g^2\chi_0^c\chi_0^n=0,
\end{equation}
which gives the line of collapse as a function of $\mu$ and $T$ (for
$T>T_c$) or as a function of $n_0$ and $T$ (for $T<T_c$).  We use the
following relations to plot the instability on the $n-T$ phase diagram,
\begin{equation}
 n = \left\{\begin{array}{ll}
    \Lambda_T^{-3} g_{3/2}(e^{\beta\mu}),&\quad T>T_c\\
    n_0 + \Lambda_T^{-3} \zeta(3/2),&\quad T<T_c.
\end{array}\right.
\end{equation}

\begin{figure}[tb]
  \epsfxsize=5in\nobreak \centerline{\epsfbox{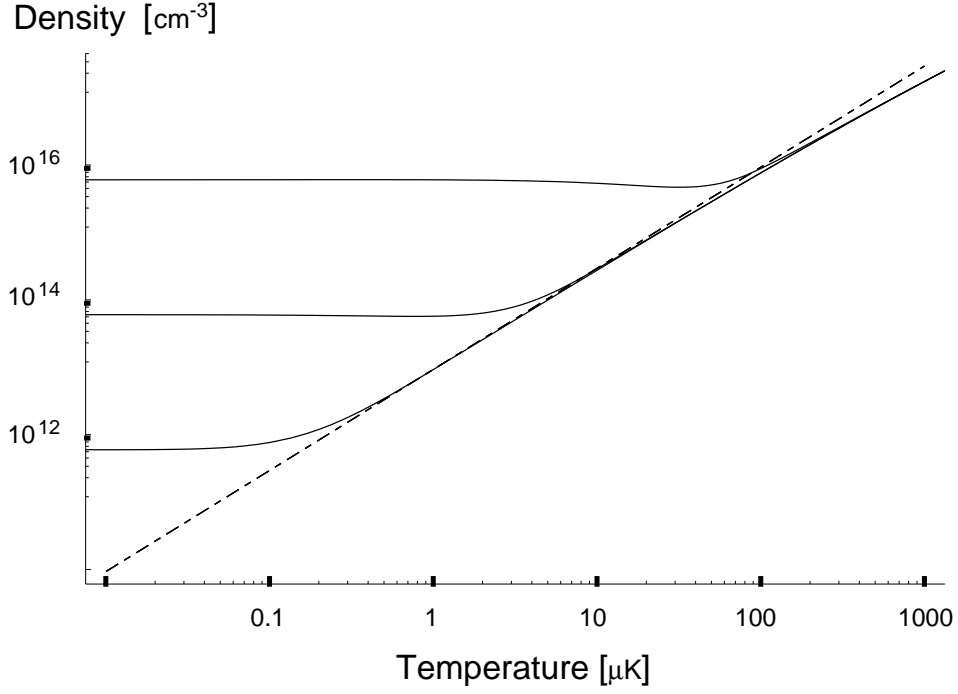}}\nobreak
  \vskip2mm\nobreak \vskip1in
\caption{
  Scaling of the instability threshold with system size.  Solid lines
  show the maximum stable density $n_{\rm max}$ for a given
  temperature.  From the top, the system size is $L$ = 0.3, 3, 30
  $\mu$m.  The other parameters are the same as in
  Fig.~\ref{FIG:phase}a).  The dashed line indicates the Bose-Einstein
  condensation transition.  Note the three scaling regimes; at low
  temperature, $n_{\rm max}\sim L^2$, near the critical temperature
  $n_{\rm max}\sim L^{3/2}$, and at high temperatures $n_{\rm max}$ is
  independent of $L$ (see Eqs.~(5) and (6)).  }\label{FIG:scaling}
\end{figure}

    Equations~(\ref{EQ:belowT}) and (\ref{EQ:belowT2}) indicate that below
$T_c$ the noncondensate response $\chi_0^n$ scales as $k^{-1}\sim L$, while
the condensate response $\chi_0^c$ scales as $k^{-2}\sim L^2$.  For realistic
parameters, $L$ is the largest length in the problem, so that the condensate
dominates the instability except when $n_0$ is much smaller than $n$.  Since
the condensate is very localized, even a few particles in the lowest mode make
$n_0$ locally much greater than the density of noncondensed particles.  In
Fig.~\ref{FIG:scaling} we show how the line of instability depends on the size
of the system, $L$.

\begin{figure}[tbh]
  \epsfxsize=5in\nobreak \centerline{\epsfbox{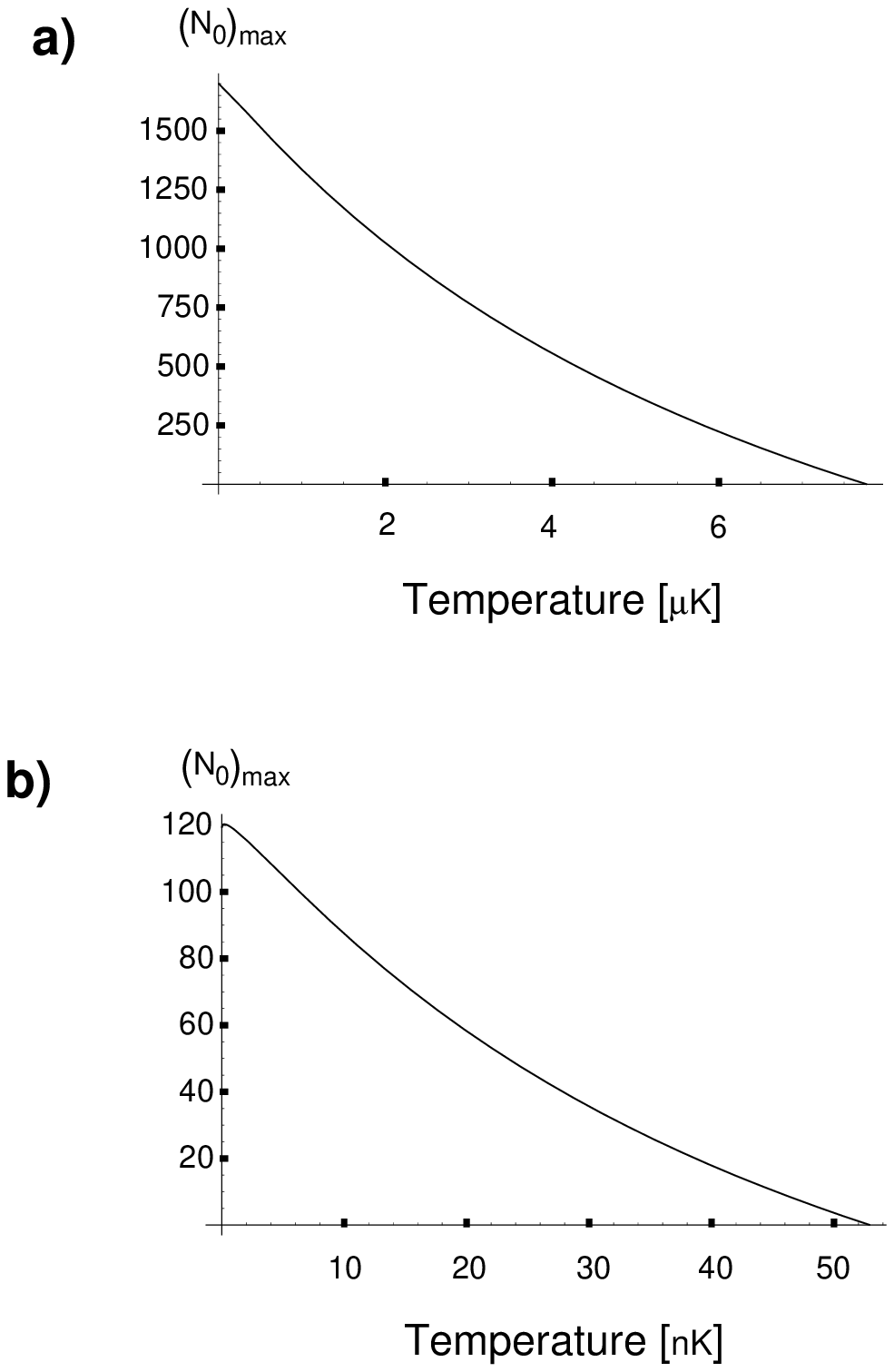}}\nobreak
  \vskip10mm\nobreak
    \caption{Maximum number of condensed particles as a function of
      temperature, for the a) $^7$Li and b) $^{85}$Rb experiments.
      The $(k\Lambda_T)^0$ terms in Eq.~\ref{EQ:belowT} have been
      included in producing these plots.  }\label{FIG:cond}
\end{figure}

    From the above equations we calculate the maximum stable value of the
condensate density $n_0$.  The line of collapse crosses the line of
condensation at a temperature $T^*=\hbar^2/2mk_BL|a|$.  Above this temperature
no condensation can occur.  For $T<T^*$ the collapse limits the density of
condensed particles to
\begin{equation}\label{EQ:max0}
(n_0)_{\rm max} = \frac{\pi}{4L^3} \left(\frac{L}{|a|}\right)
\left(\frac{T^*-T}{T^*+T} \right);
\end{equation}
$(n_0)_{\rm max}$ decreases monotonically with temperature, from the value
$(n_0)_{\rm max}^{\rm T = 0} = \pi/4L^2|a|$, eventually vanishing at $T =
T^*$.  Using parameters from experiments \cite{REF:randy,REF:JILA}, we find
for the Rice $^7$Li trap, $(N_0)_{\rm max}^{\rm T = 0} = L^3 (n_0)_{\rm
max}^{\rm T = 0} = 1700$, and $T^*$ = 7.5 $\mu$K, while for the JILA
$^{85}$Rb
trap, $(N_0)_{\rm max}^{\rm T = 0} = 120$, and $T^*$ = 46 nK, The maximum
number of particles vs. temperature for the two experiments are plotted in
Fig.~\ref{FIG:cond}; these results are consistent with the experiments, and
agree quite well with numerical mean-field calculations \cite{REF:finiteT}.
In particular, our curve $(N_0)_{\rm max}(T)$ for lithium has a slope of
$-1/2.2$ nK at $T=0$, which lies between the calculated slopes of Davis et al.
and Houbiers et al.  \cite{REF:finiteT}.  Although $(n_0)_{\rm max}$ decreases
with temperature, the non-condensed density $n^\prime$ increases ($n^\prime =
\Lambda^{-3} \zeta(3/2)$).  Thus the total density at collapse need not be
monotonic with temperature (cf. the low temperature region of
Fig.~\ref{FIG:phase}b).

Future condensate experiments at higher temperatures and densities
should be able to study the structure in Eq.~(\ref{EQ:max0}), and map
out the phase diagram in Fig.~\ref{FIG:phase}.  The rubidium
experiments are performed at temperatures near $T^*$, where the
spinodal line intersects $T_c$, and in principle should be able to
explore the crossover between the quantum mechanical and classical
behavior of the instability.  The lithium experiments are much further
away from exploring this regime, and in the current geometry,
inelastic processes make such an investigation impractical
\cite{REF:lifetime}.  Since $T^*$ is proportional to $1/L$, a softer
trap could be used to bring this crossover down to lower temperatures
where these difficulties are less severe (see Fig.~\ref{FIG:scaling}).
More precise numerical studies at higher temperatures are needed to
guide these experiments.

\section{Modeling the harmonic trap}\label{SEC:profile}

Most experimental and theoretical results are reported in terms of
numbers of particles instead of density.  By appropriately modeling
the density distribution of a harmonically trapped gas, we can present
our conclusions in such a form.  Once the interactions are strong
enough to modify the density distribution significantly, the system
undergoes collapse; thus we can take the density distribution to be
that of non-interacting particles.  For $k_B T\gg \hbar \omega$, the
density profile is well-approximated by
\begin{equation}\label{EQ:profile}
n(r) = \int\!\!\frac{d^3p}{(2\pi)^3}
\frac{1}{e^{\beta (\varepsilon_p+V(r)-\mu)}-1}
 + n_0 e^{-r^2/d^2},
\end{equation}
where $V(r)=m\omega^2r^2/2$ is the confining potential, with
characteristic length $d=(\hbar/m\omega)^{1/2}$.  The density of
condensed particles at the center of the trap is $n_0$.  Above $T_c$,
$n_0 = 0$, and below $T_c$, $\mu = 0$.  Integrating over space, we
have
\begin{equation}
N = \left\{\begin{array}{ll}
\left({k_B T}/{\hbar\omega}\right)^3 g_3(e^{\beta\mu}),
&\quad T>T_c\\
\left({k_B T}/{\hbar\omega}\right)^3 \zeta(3)+
\left(\pi\hbar/m\omega\right)^{3/2} n_0,
&\quad T<T_c.
\end{array}\right.
\end{equation}

The instability occurs in the lowest energy mode of the system, the
breathing mode, whose wave-vector is proportional to $1/d$.  In a zero
temperature non-interacting gas the breathing mode has a density
profile $\delta\rho\propto (2r^2/d^2-3) \exp{(-r^2/d^2)}$, where $r$
is the radial coordinate.  In momentum space this distribution is
peaked at wave-vector $k=2/d$.  At finite temperature thermal pressure
increases the radius of the cloud and the wave vector of the breathing
mode becomes smaller.  Since the response of the non-condensed cloud
is relatively insensitive to the wave-vector, we look for an
instability at $k=2/d$.

\begin{figure}[tb]
  \epsfxsize=5in\nobreak
  \centerline{\epsfbox{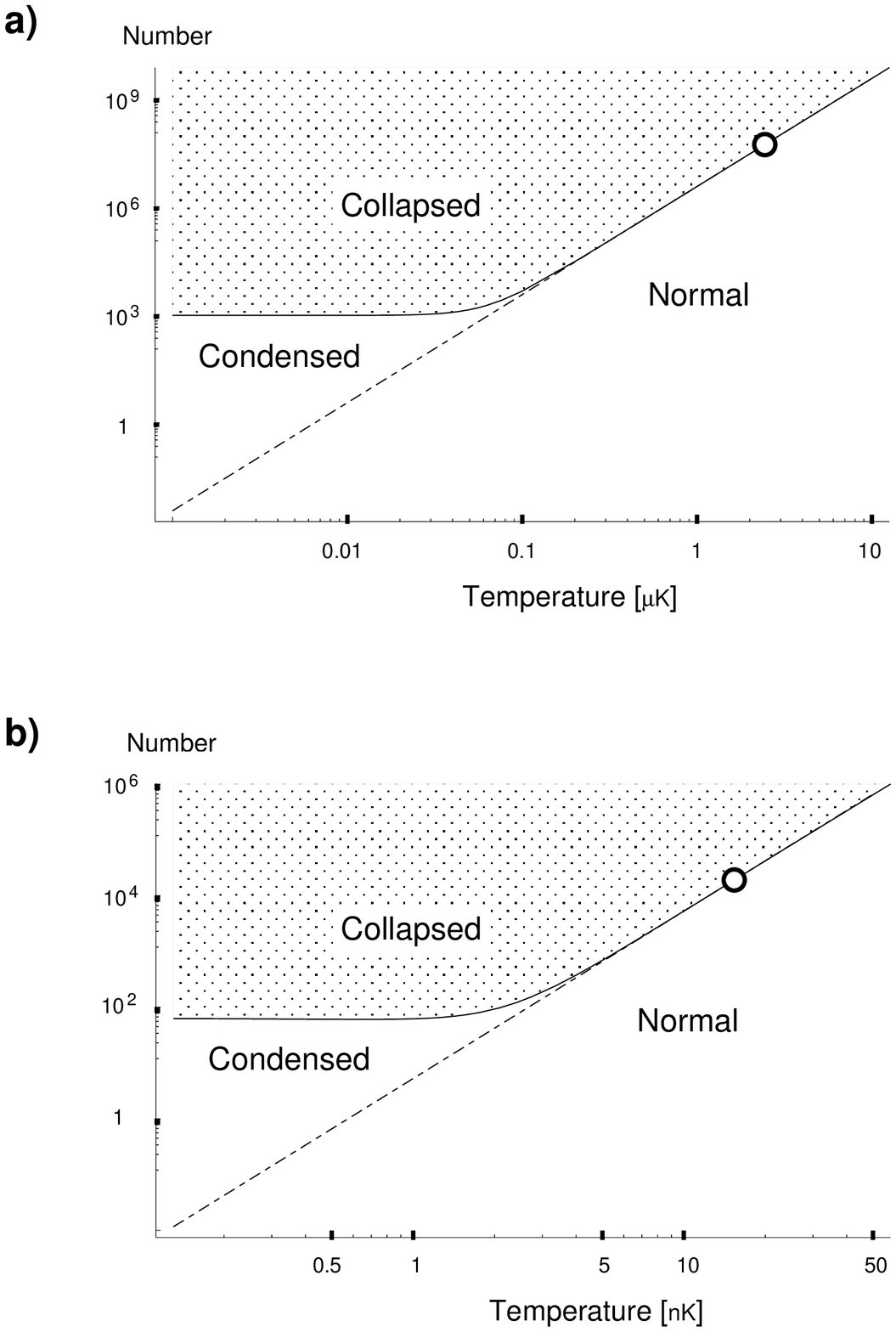}}\nobreak \vskip10mm\nobreak
 \caption{
   Phase diagram, number $N$ versus temperature $T$, for
   harmonically trapped bosons with attractive interactions;  a)
   $^7$Li, b) $^{85}$Rb.  The parameters used correspond to 
   Fig.~\ref{FIG:phase}.
   The open circle marks where the spinodal line
   meets with the BEC phase transition.
   }
\label{FIG:phase2}
\end{figure}

The resulting phase diagram, Fig.~\ref{FIG:phase2}, is similar to that
in Fig.~\ref{FIG:phase}.  The most significant difference is that the
line of collapse follows the condensation line (on the scale of the
figure they appear to coincide over a significant temperature range).
This behavior can be understood by noting that for trapped particles,
condensation results in a huge increase in the central density of the
cloud (a standard diagnostic of BEC).

\section{Pairing}\label{SEC:pairing}

With minor changes the formalism presented here can be used to
investigate the instability towards forming loosely bound dimers, or
``pairs,'' the Evans-Rashid transition.  Such an instability occurs in
an electron gas at the superconducting BCS transition \cite{REF:BCS},
and has been predicted by Houbiers and Stoof
\cite{REF:finiteT,REF:stoof} to occur in the trapped alkalis.  The
pairing is signalled by an instability in the T-matrix of the normal
phase \cite{REF:KB}, which plays the role that the density response
function plays in the collapse.  Again, we simulate the finite size of
the cloud by looking for an instability at $k=2\pi/L$, as opposed to
$k=0$ in a bulk sample.  In analogy to Eq.~(\ref{EQ:hf}), the T-matrix
can be written as a ladder sum,
\begin{equation}
{\cal T}(k,\omega) = \frac{g}{1-g \Xi(k,\omega)}.
\end{equation}
In this equation, $k$ is the relative momentum of the pair.  The
instability towards pairing is signalled by ${\cal T}\to\infty$, when
$g\Xi=1$.  To the same level of approximation as Eq.~(\ref{EQ:chio}),
the medium-dependent part of the ``pair bubble'' $\Xi$ is
\begin{equation}\label{pbubble}
\Xi(k,\omega)=
\int\frac{d^3q}{(2\pi)^3}
 \frac{ f(q-k/2)+f(q+k/2)}
      {\omega-(\varepsilon_{q+k/2} +\varepsilon_{q-k/2})}.
\end{equation}
Setting $\omega=0$, and expanding in small $k\lambda_T$, we find
\begin{equation}\label{xiexp}
g\Xi(k,\omega=0)=
-4\frac{a_s}{\Lambda_T}
    \left[\frac{4\pi}{k\Lambda_T}
          \arctan\left(
\frac{
\left|{\epsilon_k}/{4\mu}\right|^{1/2}}
{1+\left|{\epsilon_k}/{4\mu}\right|^{1/2}}
               \right)
+ g_{1/2}(e^{\beta\mu})
-\left|{\pi}/{\beta\mu}\right|^{1/2}
+{\cal O}(k\Lambda_T)
\right].
\end{equation}
Except for the argument of the arctangent, this expression is
identical to twice $g\chi_0^n$ as given in Eq.~(\ref{EQ:aboveT}).
Since arctangent is a monotonic function, and its argument here is
smaller than in Eq.~(\ref{EQ:aboveT}), we see that $g\Xi <
2g\chi_0^n$, which implies that the instability towards collapse
occurs at a lower density than the pairing instability.  Thus we
conclude that the pairing transition does not occur in an attractive
Bose gas.  Interestingly, in the classical limit, the instabilities
towards pairing and collapse coincide.

\section{Domain Formation in Spinor Condensates}\label{SEC:spins}

The approach used here to discuss the collapse of a gas with
attractive interactions also describes domain formation in spinor
condensates, and gives a qualitative understanding of experiments at
MIT \cite{REF:spin} in which optically trapped $^{23}$Na is placed in
a superposition of two spin states.  Although all interactions in this
system are repulsive, the two different spin states repel each other
more strongly than they repel themselves, resulting in an effective
attractive interaction.  The collapse discussed earlier becomes, in
this case, an instability towards phase separation and domain
formation.  The equilibrium domain structure is described in
\cite{REF:spintheory}.  Here we focus on the formation of metastable
domains.

The ground state of sodium has hyperfine spin $F=1$.  In the
experiments the system is prepared so that only the states
$|1\rangle=|F=1,m_F=1\rangle$ and $|0\rangle=|F=1,m_F=0\rangle$ enter
the dynamics.  The effective Hamiltonian is then
\begin{equation}\label{EQ:ham}
H = \int\!\!d^3r\, \frac{\nabla \psi_i^\dagger \cdot
                         \nabla \psi_i}{2m}
+ V(r)\psi_i^\dagger\psi_i + \frac{g_{ij}}{2}
 \psi^\dagger_i\psi^\dagger_j\psi_j\psi_i.
\end{equation}
where $\psi_i$ ($i=0,1$) is the particle destruction operator for the
state $|i\rangle$; summation over repeated indices is assumed.  The
effective interactions, $g_{ij}=4\pi\hbar^2 a_{ij}/m$, are related to
the scattering amplitudes $a_{F=0}$ and $a_{F=2}$, corresponding to
scattering in the singlet ($F_1+F_2=0$) and quintuplet ($F_1+F_2=2$)
channel, by \cite{REF:spin,REF:spintheory,REF:jason}:
\begin{mathletters}
\begin{eqnarray}
 \tilde a \equiv a_{11}=a_{01}=a_{10}  = a_{F=2}, \\
 \delta a \equiv   a_{11}-a_{00} = (a_{F=2}-a_{F=0})/3.
\end{eqnarray}
\end{mathletters}
Numerically, $\tilde a = 2.75$nm and $\delta a = 0.19$nm.  We
introduce $\tilde g=4\pi\hbar^2\tilde a/m$ and $\delta
g=4\pi\hbar^2\delta a/m$.  In the mean field approximation, the
interaction in Eq.~(\ref{EQ:ham}) becomes a function of $n_{m=0}$ and
n, the density of particles in the state $|0\rangle$ and the total
density, respectively:
\begin{equation}
\langle H_{int}\rangle =
\int\!\!d^3r\,\left(\frac{\tilde g}{2} n^2 - \frac{\delta g}{2}
n_{m=0}^2\right),
\end{equation}
which shows explicitly the effective attractive interaction.
Initially the condensate is static with density $n=10^{14} {\rm
  cm}^{-3}$, and all particles in state $|1\rangle$.  A
radio-frequency pulse places half the atoms in the $|0\rangle$ state
without changing the density profile.  Subsequently the two states
phase separate and form domains from 10 to 50 $\mu$m thick.  The trap
plays no role here, so we can neglect $V(r)$ in Eq.~(\ref{EQ:ham}) and
consider a uniform cloud.

Linearizing the equations of motion with an equal density of particles
in each state, we find two branches of excitations corresponding to
density and spin waves \cite{REF:gold}, \begin{mathletters}
  \begin{eqnarray} \omega_{ph}^2 &=& \left(\frac{k^2}{2 m}\right)^2 +
    \tilde{g} n \frac{k^2}{m}+ {\cal O}(\delta g),\\\label{EQ:spin}
    \omega_{sp}^2 &=& \left(\frac{k^2}{2 m}\right)^2 - \delta g\, n
    \frac{k^2}{4 m} + {\cal O}((\delta g)^2).
\end{eqnarray} \end{mathletters}
Since $\delta g>0$ spin excitations with imaginary frequencies appear.
The mode with the largest imaginary frequency grows most rapidly, and
the width of the domains formed should be comparable to the wavelength
$\lambda$ of this mode.  By minimizing Eq.~(\ref{EQ:spin}) we find
$\lambda= \sqrt{2\pi/n\,\delta a} =10 \mu$m, in rough agreement with
the observed domain size.

\section{Acknowledgements}

The authors are grateful to the Ecole Normale Sup\'erieure in Paris,
and the Aspen Center for Physics, where this work was carried out.  We
owe special thanks to Eugene Zaremba and Dan Sheehy for critical
comments.  We are particularly indebted to Henk Stoof for his
insightful recommendations and stimulating discussions, including
raising the question of the relation between the instabilities towards
pairing and collapse.  This research was supported in part by the
Canadian Natural Sciences and Engineering Research Council, and
National Science Foundation Grant No.~PHY98-00978, and facilitated by
the Cooperative Agreement between the University of Illinois at
Urbana-Champaign and the Centre National de la Recherche Scientifique.

\appendix
\section{Review of the RPA}

Here we give a brief derivation of the response function
$\chi(k,\omega)$, Eq.~(\ref{EQ:sus}), used in this paper.  Generically,
the response of a gas, $\chi=\delta n/\delta U$, is the direct
response to the perturbation $\chi_0$ plus the response to the mean
field generated by the disturbed atoms.  For a normal gas in the
Hartree approximation, $\delta n= \chi_0 \delta U + \chi_0 g \delta
n$, while including exchange gives $\delta n = \chi_0 \delta U +
\chi_0 2 g \delta n$, as in Eq.  (\ref{EQ:hf}).  The RPA amounts to
making a simple particular approximation to the polarization part
$\chi_0$.  For our purposes it suffices to take $\chi_0=\chi_0^n$, the
response of an ideal gas, Eq.~(\ref{EQ:chio}).

Generalizing the Hartree approximation to the condensed gas simply
requires replacing $\chi_0$ with the response of the non-condensed
particles $\chi_0^n$ plus the response of the condensate $\chi_0^c$,
Eq.~(\ref{EQ:chic}).

Including exchange in the degenerate gas requires some work, since
exchange only occurs in interactions involving non-condensed atoms.  A
simple technique, demonstrated by Minguzzi and Tosi
\cite{REF:minguzzi} is to look separately at the change in the density
of condensed and non-condensed atoms, $\delta n_0$ and $\delta \tilde
n$.  Within Hartree-Fock these changes are related to the applied
perturbation $\delta U$ by
\begin{equation}\label{EQ:mat}
\left(
\begin{array}{c}
\delta n_0 \\ \delta \tilde n
\end{array}\right)
= \left(\begin{array}{c}
\chi_0^c \\ \chi_0^n
\end{array}\right)
\delta U + \left(
\begin{array}{cc}
\chi_0^c & 2 \chi_0^c\\
2\chi_0^r & 2 \chi_0^r
\end{array}
\right)
\left(
\begin{array}{c}
g \delta n_0 \\ g \delta \tilde n
\end{array}\right),
\end{equation}
which gives the relationship
\begin{equation}
\chi = \frac{\delta n_0}{\delta U}+\frac{\delta \tilde n}{\delta U}=
\frac{\chi_0^c + \chi_0^n + g \chi_0^c\chi_0^n}
     {(1-g\chi^c_0)(1-2g\chi^n_0)-4g^2\chi^c_0\chi^n_0}.
\end{equation}
Diagrammatic expressions for these different approximations are shown
in Fig. \ref{FIG:hfdiagrams}.

\section{Asymptotic Expansions}

In this Appendix we derive the $k\Lambda_T\to0$ asymptotic expansions
for the functions $\chi_0^n$ and $\Xi$, defined by
Eqs.~(\ref{EQ:chio}) and (\ref{pbubble}).  These expansions are
constructed to be valid for all $\mu$.  We begin by breaking
$\chi_0^n$ into two terms, one containing $f(q-k/2)$ and one
containing $f(q+k/2)$.  After shifting $q$ by $\pm k/2$ and
integrating out the angular variables, we have
\begin{equation}
\chi_0^n(k,\omega) = \frac{m}{(2\pi)^2 k}
\int_0^\infty\!\!dq\,q^2\,f(q)
\log\left[\frac{(\bar p+k/2-q)(\bar p-k/2+q)}
{(\bar p+k/2+q)(\bar p-k/2-q)}\right],
\end{equation}
with $\bar p=m\omega/k$.  We extract the important structure by
rewriting the logarithm as an integral of the form $\int dx/x$.
Scaling all lengths by a multiple of the thermal wavelength, we find
\begin{eqnarray}
\chi_0^n(k,\omega) &=&
\frac{-m}{\pi k \Lambda_T^2}
\int_{z_-}^{z_+}
\!\! dz\,I(z)\\
I(z)&=&\int_{-\infty}^{\infty}\!\! dx\frac{x}{x-z}\frac{1}
{e^{x^2-\beta\mu}-1},
\end{eqnarray}
where $z_\pm=\omega/2\sqrt{\varepsilon_k T}\pm
k\Lambda_T/4\sqrt{\pi}$.  The integral $I(z)$ has been characterized
by Sz\'epfalusy and Kondor \cite{REF:szef}.  In particular, by
expanding the distribution function in terms of Matsubara frequencies,
one arrives at the $z\to 0$ asymptotic expansion,
\begin{eqnarray}\label{expand}
I(z)
&=& \sqrt{\pi} g_{1/2}(e^{\beta\mu}) - \frac{i\pi z}{2}
+ i \pi \left(\frac{z}{z+i\sqrt{-\beta\mu}}\right)\\\nonumber
&&- 2\pi \sum_{j=1}^\infty z^{j} \sum_{\nu=1}^\infty
\Re\left[(-\beta\mu+2\pi i \nu)^{-(j+1)/2}\right],
\end{eqnarray}
where $\Re(z)$ is the real part of $z$.  Integration of the leading
terms gives Eq. (\ref{EQ:aboveT}).

Following a similar procedure of integrating out the angular variables
we write $\Xi$ as
\begin{equation}
\Xi(k,\omega)= \frac{-2m}{\pi k \Lambda_T^2}
\int_{\bar z_-}^{\bar z_+}\!\! dz\, I(z),
\end{equation}
with $\bar z_\pm =
(k\Lambda/4\sqrt{\pi})[i\sqrt{1+2\omega/\varepsilon_k}\pm 1]$.  Using
the expansion in Eq.~(\ref{expand}) gives Eq. (\ref{xiexp}).

\begin{references}
\bibitem{REF:freq} The traps used are slightly asymmetric, and the
  frequencies quoted here are the geometric mean of the three
  frequencies along each principal axis;
  see~\cite{REF:randy,REF:JILA}.
\bibitem{REF:spinodal} In the theory of liquid-gas phase transitions,
  the spinodal line is the curve on the phase diagram where $\partial
  P/\partial V=0$, which represents the edge of the co-existence
  region, beyond which supercooled vapor cannot exist.
\bibitem{REF:lifetime} At a temperature of $5\mu$K, a $3.15\mu$m cloud
  of $^7$Li collapses at a density of $n=1.39 \times 10^{14}{\rm
    cm}^{-3}$.  At such a high density the dominant decay mechanism is
  three-body collisions, giving a lifetime $\tau = (G_3\, n^2)^{-1}$.
  Using the theoretical estimate \cite{REF:decay},
  $G_3=2.6\times10^{-28}$, we find that $\tau=200 ms$.  At $T=10\mu$K
  the lifetime is only $\tau=40 ms$.
\end{references}
\end{document}